\documentclass[useAMS,usenatbib,referee,12pt]{mn2e}
 
\newcommand{\ltsimeq}{\raisebox{-0.6ex}{$\,\stackrel
        {\raisebox{-.2ex}{$\textstyle <$}}{\sim}\,$}}
\newcommand{\gtsimeq}{\raisebox{-0.6ex}{$\,\stackrel
        {\raisebox{-.2ex}{$\textstyle >$}}{\sim}\,$}}

\begin{document}

\noindent
{\LARGE\bf The obscuration by dust of most of the growth of supermassive black holes} 

\noindent
{Alejo Mart\'\i nez-Sansigre$^{\dag}$, Steve Rawlings$^{\dag}$, Mark Lacy$^{\ddag}$, Dario Fadda$^{\ddag}$,
Francine R. Marleau$^{\ddag}$, Chris Simpson$^{\star}$, Chris J. Willott$^{\diamond}$, Matt J. Jarvis$^{\dag}$

\noindent $^{\dag}$Astrophysics, Department of Physics, University of
Oxford, Keble Road, Oxford OX1 3RH, UK

\noindent $^{\ddag}$Spitzer Science Center, California Institute of Technology, MS220-6, 1200 E. California Boulevard, Pasadena, CA 91125, USA

\noindent $^{\star}$Department of Physics, University of Durham, South Road, Durham DH1 3LE, UK

\noindent $^{\diamond}$Herzberg Institute of Astrophysics, National Research Council, 5071 West Saanich Rd, Victoria, B.C. V9E 2E7, Canada

}

\noindent\hspace*{0cm}\hrulefill

{\bf Supermassive black holes underwent periods of exponential growth
during which we seem them as quasars in the distant Universe. The
summed emission from these quasars generates the cosmic X-ray
background, the spectrum of which has been used to argue that most
black-hole growth is obscured$^{[1-2]}$. There are clear examples of
obscured black-hole growth in the form of `type-2' quasars$^{[3-5]}$,
but their numbers are fewer than expected from modelling of the X-ray
background. Here we report on the direct detection of a population of
distant type-2 quasars which is at least comparable in size to the
well-known unobscured type-1 population. We selected objects that have
mid-infrared and radio emissions characteristic of quasars, but which
are faint at near-infrared and optical wavelengths. This population is
responsible for most of the black hole growth in the young Universe
and, throughout cosmic history, black-hole growth occurs in the dusty,
gas-rich centres of active galaxies.  }

\noindent\hspace*{0cm}\hrulefill

The population of unobscured type-1 quasars is now well understood
out to redshift $z \sim 2$ by virtue of optical surveys$^{[6]}$ which provide
direct measurements around the ``break'' in the luminosity function where
the quasar population outputs most of its luminosity density. Current understanding
of the obscured type-2 quasar population is far less complete.

Examples of type-2 quasars have, for many years, been confined
mainly to radio-selected samples. Radio emission is unaffected by dust and,
 at low frequencies, 
is not strongly dependent on viewing angle so the simplest interpretation of 
optical spectroscopic follow-up of radio-selected samples is that luminous quasars 
divide roughly equally into obscured (type-2) and unobscured (type-1) objects$^{[7]}$.
However, `radio-loud' quasars are rare and may be atypical: their large-scale 
radio jets may well
modify their environments in ways which influence
whether or not the nucleus is obscured, and hence the ratio of quasars in the 
type-1 and type-2 classes$^{[8]}$.

More recently, deep X-ray surveys and optical surveys 
have also been successful in identifying 
type-2 quasars$^{[5,9,10]}$. However, optical spectroscopic follow-up 
has yet to find a type-2 population which can account fully for the X-ray background$^{[1,9,11]}$. This is 
believed to be because a substantial fraction of  type-2 quasar nuclei are hidden by
`Compton thick' material (with neutral gas column densities $N_{H} \gtsimeq 10^{28}$ m$^{-2}$) 
even when observed at high X-ray energies$^{[1,12]}$.

type-2 quasars (with visual extinction $A_{V} \gtsimeq 5$ 
magnitudes towards their nuclei) fail to outshine their host galaxy at
ultra-violet and optical wavelengths. However, at longer wavelengths  the 
obscuration becomes small so the mid-infrared  
to radio properties of type-2 quasars should be similar to those 
of their type-1 counterparts, unless the obscuration or redshift 
becomes extreme. 
Indeed a survey at 60 $\mu$m with the 
Infrared Astronomy Satellite (IRAS) 
 led to the discovery of the first high-redshift radio-quiet type-2 quasar: 
the $z = 2.3$ galaxy IRAS F10214+4724$^{[4]}$. However, 
this object was discovered only because its flux was magnified by a 
factor $\sim 50$ by a gravitational lens$^{[13]}$. From the presence of
high-excitation narrow lines and scattered broad lines$^{[14]}$
it clearly has an obscured type-2 nucleus but its X-ray emission 
seems to be obscured by Compton-thick material$^{[12]}$.

With the advent of the Spitzer Space 
Telescope$^{[15]}$, with a sensitivity $\gtsimeq 100$ times 
greater than IRAS and 
$\gtsimeq 10$ than the Infrared Space Observatory (ISO), 
type-2 quasars similar to IRAS F10214+4724 can be found from their 
mid-infrared emission without the aid of gravitational lenses. 
We chose the following selection criteria to find high-redshift type-2 quasars in the
Spitzer First Look Survey (FLS): mid-infrared 24-$\mu \rm m$ flux density
$S_{24 \mu \rm m} > 300$ $\mu$Jy; near-infrared 3.6-$\mu \rm m$ flux density
$S_{3.6\mu m} \leq 45$ $\mu$Jy; and 350~ $\mu {\rm Jy} \leq 
S_{1.4{\rm GHz}} \leq 2$ mJy, where $S_{1.4{\rm GHz}}$ is the 1.4-GHz radio
flux density.

The mid-infrared ($S_{24 \mu \rm m}$) criterion was chosen to obtain a
reliable (7-$\sigma$) catalogue from the Spitzer FLS data from the
MIPS instrument (Fadda, D., {\it et al}., in preparation). At $z = 2$
this means the quasar luminosity will be $\gtsimeq 0.2
~L^{*}_{quasar}$ (where $ L^{*}_{quasar}$ is the break luminosity
which corresponds to $M_{\rm B}= -25.7$ magnitudes$^{[16]}$); this
means we can detect sources around the break in the luminosity
function (see Supplementary Information).  Observed $24 ~\mu \rm m$
corresponds to emitted $8 ~\mu \rm m$ at $z = 2$, and dust extinction
is negligible at this wavelength unless the obscuring column is
extreme. (We adopt a $\Lambda$CDM cosmology with the following
parameters: $h = H_{0} / (100 ~ \rm km ~ s^{-1} ~ Mpc^{-1}) = 0.7$;
$\Omega_{\rm m} = 0.3$; $\Omega_{\Lambda} = 0.7$.)

The upper limit on the $3.6~\mu$m flux density (obtained using the
IRAC instrument$^{[17]}$) was chosen to select only the most distant
type-2 quasars. At $z \geq 2$ the detected 3.6-$\mu$m flux density
corresponds to light emitted at $\lambda \leq 1.2~\mu$m, so we are
able to eliminate unobscured, type-1 quasars whose flux would exceed
this limit (FIG 1). Indeed, for type-2 quasars the $S_{3.6 \mu \rm m}$
emission is likely to be dominated by starlight from the host galaxy,
allowing us to estimate `photometric redshifts' $z_{\rm phot}$ (see
Supplementary Information).  Our $S_{3.6 \mu \rm m}$ criterion
corresponds to $z_{\rm phot} \gtsimeq 1.4$ (see caption to Table 1 for
more details).

The $S_{3.6 \mu \rm m}$ criterion includes objects without IRAC detections which 
we expect to have the highest redshifts. Since the
24-$\mu$m positions are accurate to $\sim 1$ 
arcsec, the FLS radio positions$^{[18]}$,
accurate to $\sim 0.5$ arcsec, were better for spectrosocopic follow-up. However, the
main importance of the radio selection criteria is to ensure that the candidates are quasars rather
than starburst galaxies. We chose a lower limit on $S_{1.4{\rm GHz}}$ well 
above the level reached by high-redshift (submillimetre-selected) starburst galaxies without the benefit of gravitational
lensing$^{[19]}$ and an upper limit to filter out the radio-loud objects,
whose extended jets might complicate interpretation.

FIG.\ 1 shows the location of the 21 type-2 quasar candidates 
that met our selection
criteria in the $S_{24 \mu \rm m}$ versus $S_{3.6\mu m}$ plane and 
Table 1 summarises their properties. The candidates were 
found from a sky area of 3.8 deg$^{2}$ in the Spitzer FLS. No type-1 quasar ($A_{V} = 0$)
that is bright enough to 
be detected at 24 $\mu$m  will be 
faint enough at 3.6 $\mu$m to be selected (FIG.\ 1). ``Blind'' low-resolution optical 
spectroscopy$^{[20]}$ was performed on our 21 candidates, 10 of which yielded clear type-2 
spectra with narrow emission lines
giving redshifts in the range $1.4 \ltsimeq z \ltsimeq 4.2$. 
These objects are clearly type-2 quasars either because they have high-excitation 
lines (e.g.\ inset to FIG.\ 1) or because the rest-frame
equivalent widths of their Ly-$\alpha$ lines are $>100 ~ \rm nm$  
and hence significantly larger than those seen in starbursts.
Of the 11 objects that yielded no redshifts, all but one (which shows faint 
red continuum) have completely blank spectra, showing there is probably 
no contamination 
from lower-redshift ($z \ltsimeq 1.4$) starbursts since they 
would have shown [OII] line 
emission, or if highly obscured, at least some continuum.
We believe these blank-spectrum objects are also type-2 quasars 
with $z \gtsimeq 1.4$ and 
there is no compelling argument against them having $z \gtsimeq 2$ 
since the Lyman-$\alpha$ line 
can easily be extinguished if the host galaxy is dusty on large (kpc) scales. 
We are probably seeing two types of type-2: the objects with narrow emission 
lines are obscured by the torus$^{[3]}$, while the blank-spectrum objects 
are probably obscured by a starbursting host galaxy 
and we are seeing the coeval growth of 
supermassive black hole and host galaxy. Growing supermassive black holes 
embedded in a starburst 
have been found at $z \sim 2$ from X-ray measurements of 
submillimetre-selected galaxies$^{[21]}$.

To interpret our results in terms of the `quasar fraction' $q$ --
the ratio of the number of type-1 quasars to the total number of type-1 and type-2 quasars --
we need to predict the average number $<N_{1}>$ of type-1 quasars meeting identical 
$S_{24 \mu \rm m}$ and $S_{1.4{\rm GHz}}$ selection criteria, and having matched
redshift and sky area selection functions. We estimated a probability distribution p$(<N_{1}>)$
to account for the various uncertainties, the gaussian approximation of which
($4.3^{+2.2}_{-1.1}$) represents the expected number of type-1 quasars expected to adhere to both our 24-$\mu$m and 
1.4-GHz selection criteria$^{[22]}$, 
and with $z \geq 2$ in a 3.8 deg$^{2}$ patch;
this number would be 15-times higher without any radio selection criteria. 

FIG.\ 2 shows the resulting conditional probability distribution  
of the quasar fraction $q$ given our new data on type-2 quasars and 
our background knowledge of the 
type-1 quasar population, namely p$(q | {\rm data},$ $\{{\rm type-1 ~ quasar} \})$. 
This bayesian approach allows us to account properly for 
small number statistics 
(see FIG.\ 2 and Supplementary Information). 
It is difficult to judge whether the blank-spectrum objects with photometric
redshifts $z_{\rm phot}> 2$ should be included (dashed red line in FIG.\ 2) 
until we know more about the dust content of their host galaxies, but 
 considering just the objects with spectroscopic redshifts
(solid blue line in FIG.\ 2) is clearly a conservative estimate. 

The most likely systematic problem with our analysis is that we have implicitly 
assumed identical radio properties$^{[22]}$ for type-1 and 
type-2 quasars matched in intrinsic luminosity.
However, `unified models'$^{[3]}$ predict Doppler-boosted 
 radio emission from the
weak jets aligned closer to the line-of-sight for type-1 quasars$^{[23]}$ and 
hence, on average, lower radio luminosities for type-2 quasars. This would 
yield lower derived values of $q$ 
and strengthen our conclusion that most quasars are obscured. It is of 
course possible to postulate that the optical-to-radio correlation evolves 
with redshift, but this is currently an unknown 
(see also Supplementary Information).

The median redshift of our spectroscopic sample is at $z = 2$ 
which is exactly where the optical luminosity density of the 
type-1 quasar population is at its peak$^{[6]}$. Our sample shows that 
during this `epoch of quasar activity' we can be confident that about half of all
high-luminosity quasars are obscured and if some of the blank-spectrum objects turn out to
have $z \gtsimeq 2$ then it is likely that type-2 quasars outnumber type-1 quasars by a significant 
factor. It is also possible that we are missing type-2 quasars with an 
extreme obscuring column ($A_{\rm V} \gtsimeq 100$) which could be due, for 
example, to the obscuring torus being completely edge-on to the line of 
sight, but this would only strengthen our result, which 
has been predicted both by models
for the X-ray background$^{[1]}$, and by models predicting the surface 
density of
faint AGN in relatively small sky area surveys$^{[24]}$. This is, however,
 the first
direct detection of a high-redshift type-2 quasar population likely to
outnumber the type-1 population. We note that the solid blue line 
(which compares narrow-line or `torus-obscured' type-2s with type-1s) 
is consistent with the 
quasar fraction from radio-loud samples$^{[7]}$, while the dashed red line 
(which includes both `torus-' and `host-obscured' type-2s) has enough 
obscured objects to account for the X-ray background$^{[1]}$.

FIG.\ 2 shows the expected ratio for two `receding torus'$^{[25,26,27]}$ models
in which the solid angle subtended by the obscuring torus decreases (so that $q$
increases) systematically with increasing quasar luminosity.
Since we have suggested that some of the blank-spectrum objects are obscured by kpc-scale dust,
it is only the ratio of spectroscopically-confirmed type-2 quasars to
type-1 quasars (the solid blue line) that is relevant to the model predictions, 
thus these models are not excluded.

To conclude, we consider our results
alongside those of the X-ray surveys at lower redshifts$^{[9]}$ ($z \ltsimeq 2$) 
which find that most accretion at low redshift occurs in type-2 quasars and 
type-2 Seyferts (their
lower luminosity analogues). We deduce that 
throughout cosmic history, black hole growth seems to have 
been concentrated in obscured regions.
This is in good agreement with predictions from the X-ray background$^{[1]}$ 
and implies, from comparisons
between the integrated luminosity density of quasars (both type-1 and 
type-2) and 
the local space density of relic black holes$^{[28]}$, that black 
hole growth occurs in 
short, efficient spurts in the cores of forming galaxies$^{[2]}$.

\vspace*{0.3cm}
\noindent {\bf Received 24 December 2004; accepted 18 May 2005.} \newline

\vspace*{0.3cm}
\noindent\hspace*{0cm}\hrulefill 

\normalsize

\noindent
\noindent[1] Worsley, M.A., Fabian, A.C., Barcons, X., Mateos, S., Hasinger, G., The (un)resolved X-ray background in the Lockman Hole, {\it Mon. Not. R. Astr. Soc.} {\bf 354}, 720-726 (2004)  

\noindent[2] Fabian, A.C., The obscured growth of massive black holes, {\it Mon. Not. R. Astr. Soc.} {\bf 308}, L39-L43 (1999)  

\noindent[3] Antonucci, R., Unified models for active galactic nuclei and quasars, {\it Ann.Rev.Astron.Astrophys.} {\bf 31}, 473-521 (1993)  

\noindent[4] Rowan-Robinson, M., {\it et al}.  A high-redshift IRAS galaxy with huge luminosity - Hidden quasar or protogalaxy? {\it Nature} {\bf 351}, 719-721 (1991)  

\noindent[5] Norman, C., {\it et al.}, A classic Type 2 quasar, {\it Astrophys. J.} {\bf 571}, 218-225 (2002)  

\noindent[6] Wolf, C., {\it et al.}, The evolution of faint AGN between z$\sim$1 and z$\sim$5 from the COMBO-17 survey, {\it Astron. Astrophys} {\bf 408}, 499-514 (2003)  

\noindent[7] Willott, C.J., Rawlings, S., Blundell, K.M., Lacy, M., The quasar fraction in low-frequency-selected complete samples and implications for unified schemes {\it Mon. Not. R. Astr. Soc.} {\bf 316}, 449-458 (2000)  

\noindent[8] Baker, J.C., {\it et al.}, Associated absorption in radio quasars. I. C IV absorption and the growth of radio sources, {\it Astrophys. J.} {\bf 568}, 592-609 (2002)  

\noindent[9] Barger, A.J., {\it et al.}, The Cosmic Evolution of Hard X-Ray-selected Active Galactic Nuclei, {\it Astron. J.} {\bf 129}, 578-609 (2005)  

\noindent[10] Zakamska, N.L., {\it et al.}, Candidate Type II Quasars from the SDSS: III. Spectropolarimetry Reveals Hidden Type I Nuclei, {\it Astron. J.}{\bf 129}, 1212-1224 (2005)

\noindent[11] Zheng, W., {\it et al.}, Photometric redshifts of X-ray sources in the Chandra Deep Field South, {\it Astrophys. J. S.} {\bf 155}, 73-87 (2004)  

\noindent[12] Alexander, D.M., {\it et al.}, A Chandra observation of the z=2.285 galaxy FSC10214+4724: Evidence for a Compton-thick quasar? {\it Mon. Not. R. Astr. Soc.} {\bf 357}, L16-L20 (2005)

\noindent[13] Broadhurst, T., Lehar, J., A gravitational lens solution for the IRAS Galaxy FSC 10214+4724, {\it Astrophys. J.} {\bf 450}, L41-L44 (1995)  

\noindent[14] Serjeant, S., {\it et al.},  A spectroscopic study of IRAS F10214+4724, {\it Mon. Not. R. Astr. Soc.} {\bf 298}, 321-331 (1998)   

\noindent[15] Werner, M.W., {\it et al.}, The Spitzer Space Telescope Mission, {\it Astrophys. J. S.} {\bf 154}, 1-9 (2004)  

\noindent[16] Croom, S.M., {\it et al.},  The 2dF QSO Redshift Survey-XII. The spectroscopic catalogue and luminosity function, {\it Mon. Not. R. Astr. Soc.} {\bf 349}, 1397-1418 (2004)   

\noindent[17] Lacy, M., {\it et al.}, The Infrared Array Camera component of the Spitzer Space Telescope Extragalactic First Look Survey,  {\it Accepted by Astrophys. J. S.} astro-ph/0507143

\noindent[18] Condon, J.J., {\it et al.}, The SIRTF First-Look Survey. I. VLA Image and Source Catalog, {\it Astrophys. J.} {\bf 125}, 2411-2426 (2003)  

\noindent[19] Chapman, S.C., Blain, A.W., Smail, I., Ivison, R.J., A redshift survey of the submillimetre galaxy population, {\it Astrophys. J.} {\bf 622}, 772-796 (2005)

\noindent[20] Rawlings, S., Eales, S., Warren, S., The detection of four high-redshift ($0.5 \leq z \leq 3.22$) radiogalaxies by optical spectroscopy of five blank fields, {\it Mon. Not. R. Astr. Soc.} {\bf 243}, 14-18 (1990)  

\noindent[21] Alexander, D.M., {\it et al.}, Rapid Growth of Black Holes in 
Massive Star-Forming Galaxies, {\it Nature} {\bf 434}, 738-740 (2005)  

\noindent[22] Cirasuolo, M., Celotti, A., Magliocchetti, M., Danese, L., Is there a dichotomy in the radio loudness distribution of quasars? {\it Mon. Not. R. Astr. Soc.} {\bf 346}, 447-455 (2003)  

\noindent[23] Miller, P., Rawlings, S., Saunders, R., The Radio and Optical Properties of the $z<0.5$ BQS Quasars, {\it Mon. Not. R. Astr. Soc.} {\bf 263}, 425-460 (1993)  

\noindent[24] Treister, E., {\it et al.}, Obscured AGN and the X-ray, Optical and Far-Infrared Number Counts of AGN in the GOODS Fields, {\it Astrophys. J.} {\bf 616},123-135 (2004)  

\noindent[25] Lawrence, A.,  The relative frequency of broad-lined and narrow-lined active galactic nuclei -- implications for
unified schemes, {\it Mon. Not. R. Astr. Soc.} {\bf 252}, 586-592 (1991)  

\noindent[26] Simpson, C., A new look at the isotropy of narrow-line emission in extragalactic radio sources, {\it Mon. Not. R. Astr. Soc.} {\bf 297}, L39-L43 (1998)  

\noindent[27] Simpson, C., The luminosity dependence of the type 1 active galactic nucleus fraction, {\it Mon. Not. R. Astr. Soc.} {\bf 360}, 565-572 (2005)

\noindent[28] Yu, Q., Tremaine, S.,  Observational constraints on growth of massive black holes, {\it Mon. Not. R. Astr. Soc.} {\bf 335}, 965-976 (2002)  

\noindent[29] Marleau, F.R., {\it et al.}, Extragalactic Source Counts at 24 Microns in the Spitzer First Look Survey, {\it Astrophys. J. S.}, {\bf 154}, 66-69 (2004)  

\noindent[30] Bruzual, G., Charlot, S., Stellar population synthesis at the resolution of 2003, {\it Mon. Not. R. Astr. Soc.} {\bf 344}, 1000-1028 (2003)  

\noindent[31] Rowan-Robinson, M., A new model for the infrared emission of quasars, {\it Mon. Not. R. Astr. Soc.} {\bf 272}, 737-748 (1995)  

\noindent[32] Pei, Y.C., Interstellar dust from the Milky Way to the Magellanic Clouds, {\it Astrophys. J.} {\bf 395}, 130-139 (1992)

\noindent\hspace*{0cm}\hrulefill  

\noindent
{\bf Supplementary Information} is linked to the online version of the paper at  www.nature.com/nature.\newline
\noindent

{\bf Acknowledgments.}  A.M.-S. is grateful to the Council of the
European Union for financial support. S.R. and C.S.  are grateful to
the UK PPARC for a Senior Research Fellowship and an Advanced
Fellowship respectively. We thank C. Wolf, L. Clewley,
H.-R. Kl\"ockner, and G. Cotter for useful discussions, and the
referees for valuable comments. \newline
\noindent {\bf Author Information.} Reprints and permissions
information is available at npg.nature.com/reprintsandpermissions. The
authors declare no competing financial interests. Correspondence and
requests for material should be addressed to A.M.-S. (email:
a.martinez-sansigre1@physics.oxford.ac.uk).

\begin{table*}
\label{tab:sample}

\begin{center}
\begin{tabular}{lllrrrrrrl}
\hline
\hline

Name & RA & Dec & $S_{24\mu \rm m}$ &  $S_{1.4 \rm GHz}$ &  $S_{3.6\mu \rm m}$&  $S_{4.5\mu \rm m}$  &  
$z_{\rm phot}$  &  $z_{\rm spec}$  & Comments on\\

& (J2000)  & &  / $\mu $Jy & / $\mu $Jy & / $\mu $Jy &  /$\mu $Jy & & & Spectroscopy \\

\hline

AMS01 & 17 13 11.17 & +59 55 51.5  & 536 & 490 &25.0 &30.0 & 2.1 & &  Blank \\
AMS02 & 17 13 15.88 & +60 02 34.2  & 294 & 1184 & 44.5 & 61.0 & 1.4 & & Blank \\
AMS03 & 17 13 40.19 & +59 27 45.8  & 500 & 1986 & 16.0 & 28.0 & 3.1 & 2.689 & Single line$^{\dag}$ \\
AMS04 & 17 13 40.62 & +59 49 17.1  & 828 & 536 & 18.0 &22.0 &2.8 & 1.782 &\\
AMS05 & 17 13 42.77 & +59 39 20.2  & 1769 & 1038 & 34.7 & 61.4 & 1.7 & 2.022 &   Weak Lyman-$\alpha$ \\
AMS06 & 17 13 43.91 & +59 57 14.6  & 969 & 444 & $<$20 & $<$25 & $\geq$2.5 & & Blank  \\
AMS07 & 17 14 02.25 & +59 48 28.8  & 503 & 354 & 37.9 & 47.8 & 1.5 & &  Blank \\
AMS08 & 17 14 29.67 & +59 32 33.5  & 792 & 655 & 41.6 & 46.3 & 1.4 & 1.979  & \\
AMS09 & 17 14 34.87 & +58 56 46.4  & 685 & 426 & 25.2 & $<$25 & 2.1 & & Blank \\
AMS10 & 17 16 20.08 & +59 40 26.5  & 338 & 1645 & 39.2 & 44.7 & 1.5 & & Blank \\
AMS11 & 17 18 21.33 & +59 40 27.1  & 442 & 356 & 32.4 & 55.6 & 1.7 & & Blank \\
AMS12 & 17 18 22.65 & +59 01 54.3  & 518 & 946 & 25.24 & $<$25 & 2.0 & 2.770 & \\
AMS13 & 17 18 44.40 & +59 20 00.8  & 4196 & 1888 & 24.7 & 49.1 & 2.1 & 1.986 & \\
AMS14 & 17 18 45.47 & +58 51 22.5  & 937 & 469 & 8.0 & 15.0 & 4.6 & 1.504 & \\
AMS15 & 17 18 56.93 & +59 03 24.1  & 371 & 440 & 16.0 & 16.0 & 3.0 & & Blank\\
AMS16 & 17 19 42.07 & +58 47 08.9  & 788 & 390 & 18.0 & 21.0 &2.8 & 4.174 & \\
AMS17 & 17 20 45.17 & +58 52 21.3  & 1134 & 615& 10.0 & 15.0 & 3.9 & 3.138 & Single line$^{\dag}$ \\
AMS18 & 17 20 46.32 & +60 02 29.6  & 925 & 390 & $<$20 & $<$25 & $\geq$2.5 & 1.418 & \\
AMS19 & 17 20 48.00 & +59 43 20.7  & 1433 & 822 &  $<$20 & 26.9 & $\geq$2.5 & & Blank\\
AMS20 & 17 20 59.10 & +59 17 50.5  & 492 & 1268 & 45.2 & 64.9 & 1.4 & & Faint Red Continuum \\
AMS21 & 17 21 20.09 & +59 03 48.6  & 720 & 449 & 25.2 & 38.6 & 2.1 & & Blank \\

\hline
\hline

\end{tabular}
\caption{ \noindent Basic data on the 21 type-2 quasars in our sample.
The J2000.0 positions are from the Spitzer FLS radio
catalogue$^{[18]}$. RA, right ascension, Dec., declination. The MIPS
24-$\mu$m flux density $S_{24\mu \rm m}$ is obtained by point spread
function (PSF) fitting$^{[29]}$ as all objects are point sources at
the $\sim 6$ arcsecond resolution of the MIPS observations, with
positional errors $\sim 1$ arcsecond. It has a typical error of $\pm
10-15$ per cent. The 1.4-GHz flux density is the peak value (in
$\mu$Jy per beam) from the radio catalogue$^{[18]}$.  The IRAC 3.6-
and 4.5-$\mu$m flux densities are measured in 5-arcsec-diameter
apertures$^{[17]}$ and have typical errors
of $\pm 10$ per cent.  The photometric redshifts are calculated by
assuming that the 3.6-$\mu$m flux density is dominated by starlight
from a $2 L^{*}_{gal}$ galaxy ($L^{*}_{gal}$ is the break in the
galaxy luminosity function, following the models of ref.$^{[30]}$ and
passive evolution). The criterion $S_{3.6\mu \rm m} \leq 45$ $\mu$Jy
corresponds to a limiting photometric redshift $z_{\rm phot} \geq
1.4$; this was chosen to allow for scatter in the photometric redshift
estimation, whilst still filtering out type-1 quasars and low-redshift
contaminants like radio galaxies.  All of the 21 candidates were
observed spectroscopically with the dual-beam ISIS instrument at the
William Herschel Telescope, on July 2004 and April 2005.  `Blind'
low-resolution spectroscopy$^{[20]}$ was performed by offsetting from
nearby stars to the radio positions, with exposure times of $\sim 30$
minutes and a continuous wavelength coverage across the entire visible
band. Objects with a$^{\dag}$ have only one definite line in their
optical spectra, which is taken to be Ly-$\alpha$ by virtue of
extreme equivalent width, blue-absorbed line profile and lack of other
lines supportive of alternative identifications.  }

\end{center}

\end{table*}

\begin{figure*}
\begin{center}
\setlength{\unitlength}{1mm}
\begin{picture}(150,100)
\put(0,0){\includegraphics{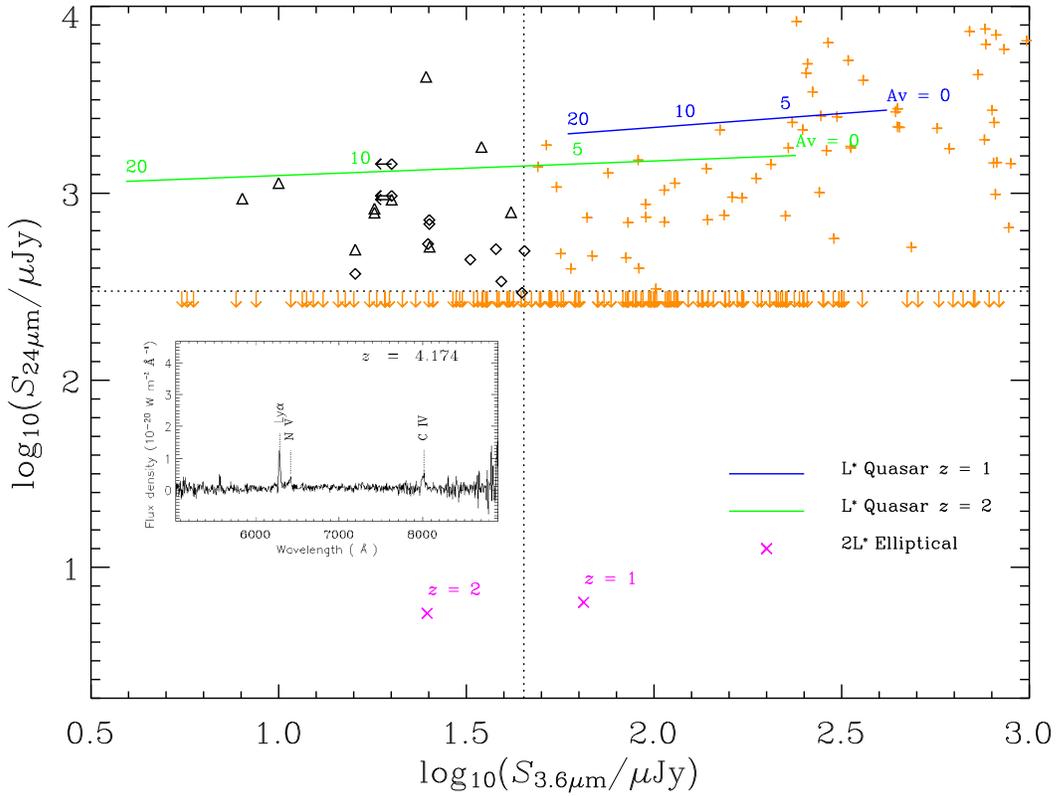}}
\end{picture}
\end{center}
{\caption{\label{fig:fig1}
Diagrammatic representation of the infrared selection 
criteria. Log$_{10}$(24 $\mu$m flux density $S_{24 \mu m}$ / $\mu$Jy) 
plotted against log$_{10}$(3.6 $\mu$m flux density $S_{3.6 \mu m}$/ $\mu$Jy). 
The dotted lines represent the flux density 
cuts of our selection criteria. Note that all objects plotted  needed 
to meet the criteria on 
radio flux density described in the text, and that the candidates were 
originally selected from the preliminary 3.6 and 24 $\mu$m catalogues, but 
have had their fluxes slightly revised. This 
explains why two candidates 
are slightly outside the selection 
boundaries (see Supplementary Information). The parent population 
is plotted as orange crosses, or arrows when $S_{24 \mu m}$ is an upper limit. 
Our 21 candidates are plotted in the top left corner, with black triangles for 
candidates with a spectroscopic redshift and black diamonds for candidates without 
one. Horizontal arrows represent upper limits for $S_{3.6 \mu m}$.  
None of the candidates showed 
type-1 quasar spectra.
The spectrum of the $z = 4.174$ object is shown in the inset. 
The magenta crosses show  $S_{24 \mu \rm m}$ and $S_{3.6 \mu \rm m}$ 
for a $2L^{*}_{gal}$ elliptical galaxy at  $z = 1$ and $z = 2$; 
starlight is assumed to evolve passively following 
a stellar population synthesis model$^{[30]}$.
The other  colours represent an $L^{*}_{quasar}$ quasar (spectral energy 
distribution from ref$^{[31]}$) with varying amounts of visual extinction 
$A_{V}$ applied to $S_{24 \mu \rm m}$ and $S_{3.6 \mu \rm m}$ and plotted on 
the curves (assuming a Milky-Way-like 
extinction curve from ref.$^{[32]}$); these are plotted at two different 
redshifts,  and the quasar population is assumed to
undergo pure luminosity evolution$^{[16]}$. At all redshifts, 
any type-1 ($A_{V} = 0$) quasar 
bright enough to make 
the  $S_{24 \mu \rm m} > 300 ~\mu$Jy criterion will also be too bright at 
3.6-$\mu \rm m$ to make it into  our sample. Inset, the optical spectrum of AMS16, showing Lyman-$\alpha$, N V and C IV emission lines.
} }
\end{figure*}

\begin{figure*}
\begin{center}
\setlength{\unitlength}{1mm}
\begin{picture}(150,100)
\put(0,0){\includegraphics{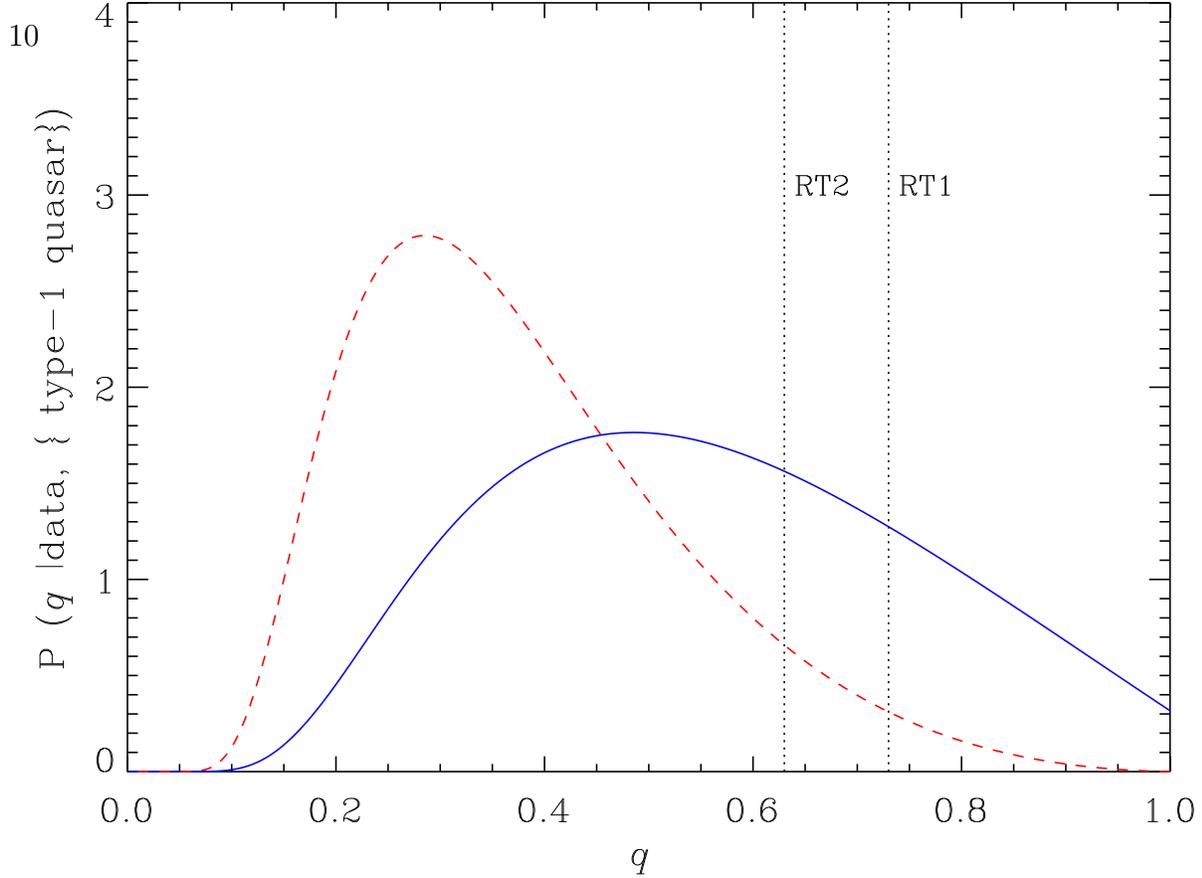}}
\end{picture}
\end{center}
{\caption{\label{fig:fig2}
Probability distributions for the quasar fraction. The 
normalised posterior probability p$(q | {\rm data, \{type-1 ~ quasar\}})$
of the quasar fraction $q$ given our new data on type-2 quasars, and 
background information
on type-1 quasars. The two lines represent different interpretations of the
number of type-2 quasars at $z \geq 2$ determined from our new data.
To obtain p$(q | {\rm data, \{type-1 ~ quasar\}})$ we used Bayes' theorem
with a uniform prior over the range $0 \leq q \leq 1$ so that the
p$(q | {\rm data, \{type-1 ~ quasar\}}) = {\rm p}({\rm data} | q,\{\rm{type-1 ~ quasar}\})$. 
We then calculated this
likelihood function, at each $q$, as a Poisson distribution for the 
observed (integer) 
number of type-2 quasars
at $z \geq 2$ where the mean number expected $<N_{2}> = (1 - q) <N_{1}> / q$. 
Because of uncertainties
in the background information on $<N_{1}>$, we had to evaluate the likelihood at 
each $q$ and $<N_{1}>$ and then
integrate p$<N_{1}> \times$ the likelihood over $<N_{1}>$.
The solid blue line is the posterior probability if we believe the only candidates with $z \geq 2$ 
are the 5 whose spectroscopic redshifts confirm this. If, additionally, 
we use the photometric redshifts of the candidates 
with blank spectra (6 additional candidates with $z \geq 2$), 
then  we obtain the red curve. 
The best estimate for $q$ is probably somewhere 
between the 
two curves.
The two vertical lines represent the predictions of 
receding torus models, one with a fixed torus height (marked `RT1')$^{[26]}$ and one (`RT2') 
with a torus height
varying with luminosity$^{[27]}$.
}}

\end{figure*}

\end{document}